\documentclass[aps,groupedaddress,twocolumn]{revtex4}
\usepackage{epsf}
\usepackage{epsfig}
\begin{document}

\title {
\small submitted to PRL\\[0.2cm]
\large
A measurement of the evolution of Interatomic Coulombic Decay in the time domain}

\author{F. Trinter$^{1}$}
\author{J.~B.~Williams$^{1}$}
\author{M. Weller$^{1}$}
\author{M. Waitz$^{1}$}
\author{M. Pitzer$^{1}$}
\author{J. Voigtsberger$^{1}$}
\author{C. Schober$^{1}$}
\author{G. Kastirke$^{1}$}
\author{C. M\"uller$^{1}$}
\author{C. Goihl$^{1}$}
\author{P. Burzynski$^{1}$}
\author{F. Wiegandt$^{1}$}l
\author{T. Bauer$^{1}$}
\author{R. Wallauer$^{1}$}
\author{H. Sann$^{1}$}
\author{A. Kalinin$^{1}$}
\author{L. Ph. H. Schmidt$^{1}$}
\author{M. Sch\"offler$^{1}$}
\author{N. Sisourat$^{2}$}
\author{T. Jahnke$^{1}$}
\email{jahnke@atom.uni-frankfurt.de}

\affiliation{
$^1$ Institut f\"ur Kernphysik, Goethe Universit\"at, Max-von-Laue-Str.1, 60438 Frankfurt, Germany \\
$^2$ Universit\'{e} Pierre et Marie Curie, UMR 7614, Laboratoire de Chimie Physique Mati\`{e}re et Rayonnement, 11 rue Pierre et Marie Curie, F-75005 Paris, France\\
}

\date{\today}

\begin{abstract}

During the last 15 years a novel decay mechanism
of excited atoms has been discovered and
investigated. This so called ''Interatomic
Coulombic Decay'' (ICD) involves the chemical
environment of the electronically excited atom: the excitation energy is transferred (in many cases over long distances) to a neighbor
of the initially excited particle usually
ionizing that neighbor. It turned out that ICD is
a very common decay route in nature as it occurs
across van-der-Waals and hydrogen bonds. The time
evolution of ICD is predicted to be highly
complex, as its efficiency strongly depends on
the distance of the atoms involved and this
distance typically changes during the decay. Here
we present the first direct measurement of the
temporal evolution of ICD using a novel
experimental approach.

\end{abstract}
\maketitle

In 1997 Cederbaum and coworkers realized that the
presence of loosely bound atomic or molecular
neighbors opens a new relaxation pathway to an
electronically excited atom or molecule. In the
decay mechanism they proposed - termed
Intermolecular Coulombic Decay (ICD) - the
excited particle relaxes efficiently by
transferring its excitation energy to a
neighboring atom or molecule
\cite{Cederbaum97prl}. As a consequence the atom
or molecule receiving the energy emits an electron of low
kinetic energy. The occurrence of ICD was proven
in experiments in the mid 2000s by means of
electron spectroscopy \cite{Marburger03prl} and
multi-coincidence techniques \cite{Jahnke04bprl}.
Since that time a wealth of experimental and
theoretical studies have shown that ICD is a
rather common decay path in nature, as it occurs
almost everywhere in loosely bound matter. It has
been proven to occur after a manifold of initial
excitation schemes such as innervalence shell
ionization, after Auger
cascades \cite{Morishita06prl,Ueda_JSERP08},
resonant excitation \cite{Barth_JCP05,Aoto_PRL06}, shakeup
ionization \cite{Jahnke07} and resonant Auger
decay. ICD has also been observed in many systems as rare gas
clusters \cite{Ohrwall04}, even on surfaces \cite{Grieves11prl} and small water
droplets \cite{Jahnke10NPhys,Mucke10NPhys}. The
latter suggested that ICD might play a role in
radiation damage of living tissue
\cite{Kim11PNAS}, as it creates low energy
electrons, which are known to be
genotoxic \cite{Boudaiffa00sci,Hanel03prl}. More
recently that scenario was reversed as it was
suggested to employ ICD in treatment of tagged
malignant cells \cite{Trinter13Nat}. Apart from these potential
applications the elementary process of ICD is
under investigation, as the decay is predicted to
have a highly complex temporal behavior. The
efficiency and thus the decay times of ICD depend
strongly on the size of the system, i.e. the
number of neighboring particles and the distance
between them and the excited particle.
However, even for most simple possible model systems
consisting of only two atoms the temporal
evolution of the decay is non-trivial and predicted theoretically to exhibit exciting physics \cite{Kuleff07prl}: as ICD
happens on a timescale that is fast compared to
relaxation via photon emission, but comparable to
the typical times of nuclear motion in the
system, the dynamics of the decay is complicated
and so far only theoretically explored. As the
decay rates strongly depend on the internuclear
distances of the atoms participating in the decay
the correct description of the nuclear motion as
well as the precise decay widths for each
distance are both vital even for predicting
relatively general quantities, such as the energy
spectrum of the emitted ICD electrons. Examining
the temporal evolution of ICD in an experiment is
therefore one of the grand challenges in
ultrafast science. Here we present an
experimental study resolving ICD in a helium
molecule, a so called helium dimer (He$_2$), in
the time domain.

\begin{figure}[htbp]
 \begin{center}
  \epsfig{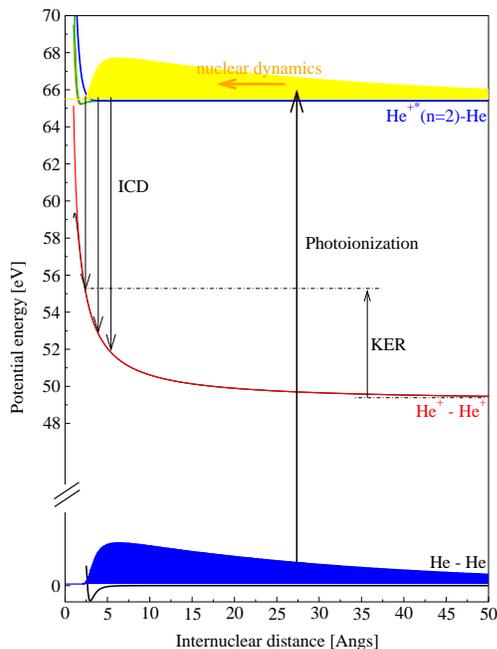}
 \end{center}
 \vspace*{-8mm}
\caption{Sketch of the potential energy diagram of the states involved in the process. The groundstate which is only bound by 95~neV is photoionized and excited. As the mean internuclear distance of the excited state is much smaller than that of the groundstate, nuclear motion sets in: the vibrational wavepacket starts to evolve on the potential energy curves of the excited states. During that time ICD happens mapping the evolving vibrational wavepacket to the repulsive He$^+/$He$^+$ final states. By measuring the kinetic energy release (KER) information on the internuclear distance (i.e. the distribution of the wavepacket) at the instant of the decay is obtained.}
 \label{pot}
\end{figure}

\begin{figure}[htbp]
 \begin{center}
  \epsfig{file=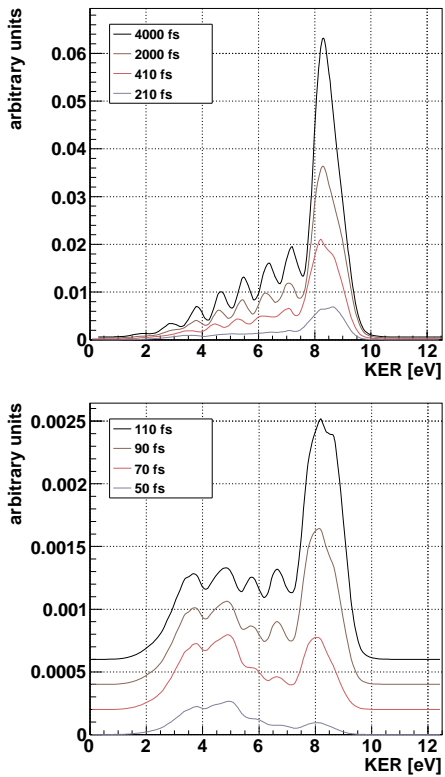,width=6.5cm}
 \end{center}
 \vspace*{-8mm}
\caption{Theoretical predictions of the time evolution of the kinetic energy release for different decay times. Bottom to top: KER for a time integrated from 0~fs to 50~fs, 70~fs, 90~fs, 110~fs, 210~fs, 410~fs, 2000~fs and to 4000~fs.}
 \label{fig1}
\end{figure}

The helium dimer is known to be the most weakly
bound ground state system in the universe \cite{Schollkopf94Science}
with a binding energy of only 95~neV (1.1~mK) and a bond length that extends
from about 5~\AA~over its mean value of 52~\AA~into the
macroscopic regime of a few hundred angstroms. Nonetheless
even in this extended system ICD occurs
transferring about 40~eV of energy from one
helium atom to its neighbor. The existence of ICD
in the helium dimer has been shown in
\cite{Havermeier10prl}. While initially ICD was
investigated after innervalence ionization in the
case of helium simultaneous photoionization and
excitation was used to produce an intermediate
ionic dimer state that is able to undergo ICD. A
multi-coincidence measurement yielded not only
the proof of the existence of ICD even in a
system as extended as the helium dimer, but for
the first time showed the occurrence of nodal
structures in the measured energy distributions
\cite{Sisourat10NatPhys}. Previously expected for
the neon dimer \cite{Scheit03jcp} these occur as the
vibrational wavefunction of the excited
intermediate dimer state is mapped onto the
repulsive final state after ICD visualizing
directly the wave-nature of the vibrating nuclei.
A sketch of the process and the involved potential energy curves is shown in Fig. \ref{pot}.
A key feature to ICD in He$_2$ is the long
distance over which the energy transfer takes
place. Consequently the decay times here are in
comparison to that of other systems very long,
allowing for the vibrational structure to form in
the excited state.
Following ICD in the time domain
thus equals following the evolution of the
vibrational wavepacket in time, as it is
triggered at large internuclear distances and then
evolves towards showing the vibrational features
shown in Fig. \ref{fig1}.

We observe this time evolution, which takes place on a
femto- to picosecond timescale making use of a new experimental
technique, which maps time to kinetic energy of an emitted electron. Such a
mapping of time to energy is typically employed in
attosecond science by streaking of electrons with
a time varying external field \cite{Drescher02Nat}. In our
novel technique the time dependent field is
created by the decaying system itself and the
photoelectron, which we launch in the pump step,
acts as the probe particle, which experiences the
streaking. In the present case we therefore used Cold Target Recoil Ion
Momentum Spectroscopy (COLTRIMS)
\cite{doerner00pr,ullrich03rep,Jahnke04JESRP} to
measure the energy of the
photoelectron carrying the time information and
the fragment ions on which we observe the time
evolution of ICD in coincidence. By expanding helium gas at a pressure of 8~bar through a nozzle with a diameter of 5$~\rm{\mu}$m and cooling the nozzle to a temperature of 22~K a supersonic gas jet containing a mixture of helium monomers and dimers was created. At these conditions a mixture of dimers and trimers might occur. The measurements, however, show that the contribution of trimers was small, as the measured spectra are expected to differ drastically for dimers and trimers. The supersonic jet was intersected by a linearly polarized  photon beam with a photon energy of $h\nu = 65.536~$eV at beamline UE112-PGM-1 at the BESSY synchrotron facility. The ion detection covered full solid angle of emission for kinetic energy releases up to 3~eV. Ions with higher kinetic energy are detected depending on their emission angle, i.e. ions being emitted within a small cone along the spectrometer axis are detected over the complete range of KERs occurring in the reaction. These events have been used as a reference for a solid angle correction of the measured KER spectra.

\begin{figure}[htbp]
 \begin{center}
  \epsfig{file=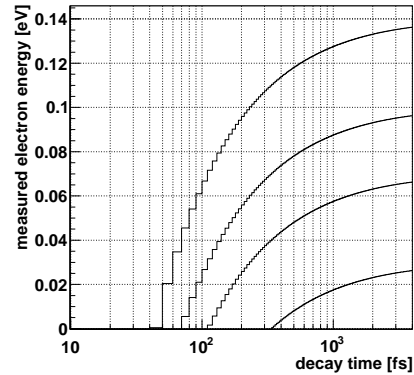,width=6.5cm}
 \end{center}
 \vspace*{-8mm}
\caption{Dependence of the shift in electron energy and the decay time obtained from our classical model. The plot depicts on the y-axis the energy a measured electron will have if the decay happens after a certain time (shown on the x-axis). The behavior is plotted for different initial photoelectron energies. From bottom to top: 30~meV, 70~meV, 100~meV, 140~meV.}
 \label{t_model}
\end{figure}

Even though it is rarely stressed in literature, the mapping of decay time to photoelectron energy
naturally occurs whenever a decay produces a
secondary electron, which is significantly faster
than the photoelectron \cite{Schuette12prl,Bauch12pra}. The change of the kinetic energy of the emitted particles is known as ''post collision
interaction'' (PCI) \cite{Niehaus77jpb}. So far, PCI has been studied
in great detail after Auger decay \cite{Sheinerman06jpb}. As the
Auger electron is emitted in the decay, the
charge of the remaining ion changes. Accordingly
an emerging photoelectron starts to leave a
singly ionized atom, but as the decay happens,
the photoelectron is suddenly exposed to the
Coulomb force of a doubly charged ion. This
results in lowering the energy of the emerging
photoelectron and increasing the kinetic energy
of the Auger electron. The energy shift of the
photoelectron wavepacket depends on the time the
Auger electron needs to emerge from the ion: a
strong shift can be expected if the Auger electron is gone
instantly, as the photoelectron is still close
to the ion and the difference between the singly and
the doubly charged potential is large in that
case. The shift decreases for longer decay times
correspondingly. Therefore, as the shift of the
electron energy can be measured, a way to access
the time domain of an electronic decay in an
experiment arises. The only requirement for this
scenario to work is, that the photoelectron is
much slower than the secondary electron. Thus by
tuning the photoelectron to very low energies the
scheme can be used to measure the evolution of ICD.
In order to convert the measured shift in energy of the photoelectron into a decay time we used a simple classical model. In a simulation an electron of a kinetic energy of 140~meV is launched. A second electron (the ICD electron) with a kinetic energy of 10~eV is launched after a delay time $t_{ICD}$. As the ICD electron reaches the photoelectron the distance the photoelectron travelled $R_p$ is obtained. The energy difference between a Coulomb potential of charge two and a Coulomb potential of charge one at $R_p$ is the amount of energy the electron is decelerated. This most simple model already shows a strong non-linear behavior for the dependency of the emission time of the second electron and the energy shift the first electron experiences as shown in Fig. \ref{t_model} for different initial (i.e. unshifted) energies of the photoelectron. Apart from being a fully classical model it furthermore neglects effects that occur due to the different emission angles of the two electrons. However, this effect is known to be strong only for a small region of almost equal emission directions \cite{Landers09prl}. It furthermore turns out,
that the minimum time that can be investigated
depends on the initial energy of the photoelectron. This is basically due to the fact, that
electrons that exhibit a severe shift are
recaptured into the ion. Therefore, choosing
an unperturbated energy of 140~meV for the photoelectron yields a minimum accessible decay time
of ~50~fs within our simple model. In the
experiment this was implemented by employing a
photon energy of 65.536~eV. At this energy one
helium atom of the dimer is ionized and excited
to (n$=2$) and emits a photoelectron of an energy of 140~meV. This excited
state can undergo ICD and was used (at higher
photon energies) in the past to identify ICD in
He$_2$ \cite{Havermeier10prl}.

In the present case the temporal evolution of the
kinetic energy release (KER) is investigated. The
KER is the energy that the two nuclei gain after
dissociating in a Coulomb explosion as ICD
occurred. The KER closely corresponds to the
internuclear distance of the two atoms of the
dimer, at the instant they were ionized: within the so called ''reflection
approximation'' \cite{Gislason73jcp} the Coulomb interaction
yields (in atomic units) the following
simple relation: $\rm{KER} = 1/R$. The results from
the theoretical investigation shown in Fig.
\ref{fig1} depict the KER for different times at
which ICD happened. At short times a first peak
at lower kinetic energies occurs. This can be
understood classically: as the internuclear
distance of He$_2$ in the groundstate is much
larger than in the excited state the decay starts
to evolve at larger internuclear distances, i.e.
smaller KERs. After some time the main peak at
high KERs builds up as the dimer contracts
towards the mean internuclear distance of the
excited ionic state. As this happens, the
probability for ICD increases (which is
proportional to $1/R^6$ at large distances \cite{averbukh04prl}) as
Fig. \ref{fig1} reveals. At longest times finally
the vibrational features form, yielding the
distribution, which is known from the
non-timeresolved investigation
\cite{Havermeier10prl,Sisourat10NatPhys}. The time resolved KER spectra, shown in Fig. \ref{fig1}, were computed using the approach reported in \cite{Chiang11prl}. The electronic structure input data used for these computations are presented and discussed in \cite{Kolorenc10pra}.

\begin{figure}[htbp]
 \begin{center}
  \epsfig{file=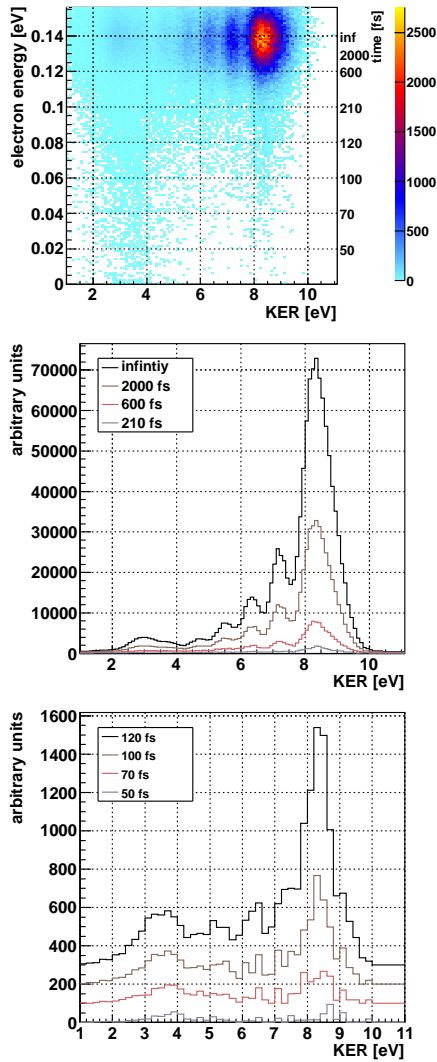,width=6.0cm}
 \vspace*{-8mm}
 \end{center}
\caption{Experimental results. Top: electron energies and kinetic energy releases measured in coincidence. The axis depicting the decay time is obtained according to our model described in the methods section. Middle and bottom: Measured kinetic energy releases corresponding to an integration over times from zero to (top to bottom): ''infinity'', 2000~fs, 600~fs, 210~fs, 120~fs, 100~fs, 70~fs, and 50~fs. The scaling of the y-axis is in both plots the same. All results shown here are solid angle corrected spectra (see methods for details), i.e. for KERs below 3.3 eV the y-axis shows counts, while the data with KER$>3.3$~eV is multiplied by a KER dependent factor up to 9 for a KER of 12~eV.}
 \label{fig2}
\end{figure}

In Fig. \ref{fig2} the experimental results are
depicted. The top panel shows the correlation of
the measured KER and the kinetic energy of the
electrons. As expected the electron spectrum
consists of a photoelectron line at an energy of
approx. 140~meV that is streaked towards lower electron
energies. The plot reveals the expected behavior:
at lowest photoelectron energies (which
correspond to shortest ICD times) mainly low
KERs occur. As the electron energy increases the
main peak at a KER of about 8.5~eV builds up. For
even later times the vibrational structures form.
The two lower panels of Fig. \ref{fig2} show the
KER for different slices in the electron energy.
The maximum photoelectron energy increases from
bottom to top, accordingly the investigated time
interval increases. The bottom panel shows the
KER for electron energies of 0~meV to 20~meV,
40meV,~60~meV, and 80~meV. The middle panel
depicts KERs from 0~meV to 100~meV, 120~meV,
135~meV and 160~meV (full range), i.e. according to our model
the time domain from 50~fs to ''infinity''. The
experimental results furthermore confirm the
findings of Fig. \ref{fig1}, that the decay
times of ICD in He$_2$ are (due to the dimer's
dimension) in the range of a few
100~femtoseconds to picoseconds, as the main
contribution to the gathered data occurs at these
times.

In conclusion we have added a new powerful
streaking approach to the toolbox of ultrafast
science and applied it to visualize the time
dependence of an interatomic decay process. The
results directly show the evolution of the
vibrational wavepacket of a helium dimer during
the decay and thus give insight into the complex
behavior of ICD in the time domain. The
measurement approach presented here can be used
to investigate other processes and systems in the
time domain, as well. Experiments investigating
the evolution of a hole created inside an atom or
molecule and for example the hopping of core
holes in molecules could be traced in time in the
future using the same approach.

\section{Acknowledgments}
R.W. and T.J. would like to thank the Deutsche Forschungsgemeinschaft (DFG) for financial support. This research has been performed within the DFG-Forschergruppe FOR1789. We acknowledge the support from the staff at BESSY during the beamtime. We are indebted to R. D\"orner for proposing this experiment.

\bibliographystyle{unsrt}

\begin{thebibliography}{10}


\bibitem{Cederbaum97prl}
L.~S. Cederbaum, J.~Zobeley, and F.~Tarantelli.
\newblock Giant intermolecular decay and fragmentation of clusters.
\newblock {\em Phys. Rev. Lett.}, \textbf{79}, 4778 (1997).

\bibitem{Marburger03prl}
S.~Marburger, O.~Kugeler, U.~Hergenhahn, and T.~M\"oller.
\newblock Experimental evidence for Interatomic Coulombic decay in Ne clusters.
\newblock {\em Phys. Rev. Lett.}, \textbf{90}, 203401 (2003).

\bibitem{Jahnke04bprl}
T. Jahnke, A. Czasch, M. S. Sch\"offler, S. Sch\"ossler, A. Knapp,
M. K\"asz, J. Titze, C. Wimmer, K. Kreidi, R. E. Grisenti, A.
Staudte, O. Jagutzki, U. Hergenhahn, H. Schmidt-B\"ocking, and R.
D\"orner.
\newblock Experimental Observation of Interatomic Coulombic Decay in Neon Dimers.
\newblock {\em Phys. Rev. Lett.}, \textbf{93}, 163401 (2004).

\bibitem{Morishita06prl}
Y. Morishita, X.-J. Liu, N. Saito, T. Lischke, M. Kato, G. Pr\"umper, M. Oura, H. Yamaoka, Y. Tamenori, I.H. Suzuki, and K. Ueda.
\newblock Experimental Evidence of Interatomic Coulombic Decay from the Auger Final States in Argon Dimers.
\newblock {\em Phys. Rev. Lett.}, \textbf{96}, 243402 (2006).

\bibitem{Ueda_JSERP08}
K. Ueda, H. Fukuzawa, X.-J. Liu, K. Sakai, G. Pr\"umper, Y.
Morishita, N. Saito, I.H. Suzuki, K. Nagaya, H. Iwayama, M. Yao,
K. Kreidi, M. Sch\"offler, T. Jahnke, S. Sch\"ossler, R. D\"orner,
Th.Weber, J. Harries, and Y. Tamenori.
\newblock Interatomic Coulombic decay following the Auger decay: Experimental evidence in rare-gas dimers.
\newblock {\em Journal of Electron Spectroscopy and Related
Phenomena}, \textbf{3-10}, 166-167 (2008).

\bibitem{Barth_JCP05}
S. Barth, S. Joshi, S. Marburger, V. Ulrich, A. Lindblad, G.
\"Ohrwall, O. Bj\"orneholm, and U. Hergenhahn.
\newblock Observation of resonant Interatomic Coulombic Decay in Ne clusters.
\newblock {\em J. Chem. Phys.}, \textbf{122}, 241102 (2005).

\bibitem{Aoto_PRL06}
T. Aoto, K. Ito, Y. Hikosaka, E. Shigemasa, F. Penent, and P.
Lablanquie.
\newblock Properties of Resonant Interatomic Coulombic Decay in Ne dimers.
\newblock {\em Phys. Rev. Lett.}, \textbf{97}, 243401 (2006).

\bibitem{Jahnke07}
T.~Jahnke A.~Czasch, M.~Sch\"offler, S. Sch\"ossler, M. K\"asz, J.
Titze, K. Kreidi, R. E. Grisenti, A. Staudte, O. Jagutzki, L. Ph.
H. Schmidt, Th. Weber, H. Schmidt-B\"ocking, K. Ueda, and R. D\"orner.
\newblock Experimental Separation of Virtual Photon Exchange and Electron Transfer in Interatomic Coulombic Decay of Neon Dimers.
\newblock {\em Phys. Rev. Lett.}, \textbf{99}, 153401 (2007).

\bibitem{Ohrwall04}
G. \"Ohrwall, M. Tchaplyguine, M. Lundwall, R. Feifel, H. Bergersen, T. Rander, A. Lindblad, J. Schulz, S. Peredkov, S. Barth, S. Marburger, U. Hergenhahn, S. Svensson, and O. Bj\"orneholm.
\newblock Femtosecond Interatomic Coulombic Decay in Free Neon Clusters: Large Lifetime Differences between Surface and Bulk.
\newblock {\em Phys. Rev. Lett.}, \textbf{93}, 173401 (2004).

\bibitem{Grieves11prl}
G. A. Grieves, and T. M. Orlando.
\newblock Intermolecular Coulomb Decay at Weakly Coupled Heterogeneous Interfaces.
\newblock {\em Phys. Rev. Lett.}, \textbf{107}, 016104 (2011).

\bibitem{Jahnke10NPhys}
T. Jahnke, H. Sann, T. Havermeier, K. Kreidi, C. Stuck, M. Meckel, M. Sch\"offler, N. Neumann, R. Wallauer, S. Voss, A. Czasch, O. Jagutzki, A. Malakzadeh, F. Afaneh, Th. Weber, H. Schmidt-B\"ocking, and R. D\"orner.
\newblock Ultrafast Energy Transfer between Water Molecules.
\newblock {\em Nature Physics}, \textbf{6}, 139 (2010).

\bibitem{Mucke10NPhys}
M. Mucke, M. Braune, S. Barth, M. F\"orstel, T. Lischke, V. Ulrich, T. Arion, U. Becker, A. Bradshaw, and U. Hergenhahn.
\newblock A hitherto unrecognized source of low-energy electrons in water.
\newblock {\em Nature Physics}, \textbf{6}, 143 (2010).

\bibitem{Kim11PNAS}
H.-K. Kim, J. Titze, M. Sch\"offler, F. Trinter, M. Waitz, J. Voigtsberger, H. Sann, M. Meckel, C. Stuck, U. Lenz, M. Odenweller, N. Neumann, S. Sch\"ossler, K. Ullmann-Pfleger, B. Ulrich, R. Costa Fraga, N. Petridis, D. Metz, A. Jung, R. Grisenti, A. Czasch, O. Jagutzki, L. Schmidt, T. Jahnke, H. Schmidt-B\"ocking, and R. D\"orner.
\newblock Enhanced production of low energy electrons by alpha particle impact.
\newblock {\em PNAS}, \textbf{108}, 11821 (2011).

\bibitem{Boudaiffa00sci}
B. Boudaiffa, P. Cloutier, D. Hunting, M. A. Huels, and L. Sanche.
\newblock Resonant formation of DNA strand breaks by low-energy (3 to 20 eV) electrons.
\newblock {\em SCIENCE}, \textbf{287}, 1658 (2000).

\bibitem{Hanel03prl}
G. Hanel, Gstir B., Denifl S., Scheier P., Probst M.,
Farizon B., Farizon M., Illenberger E., and Mark~T. D.
\newblock Electron attachment to uracil: Effective destruction at subexcitation energies.
\newblock {\em Phys. Rev. Lett.}, \textbf{90}, 188104 (2003).

\bibitem{Trinter13Nat}
F. Trinter, M. S. Sch\"offler, H.-K. Kim, F. Sturm, K. Cole, N. Neumann, A. Vredenborg, J. Williams, I. Bocharova, R. Guillemin, M. Simon, A. Belkacem, A. L. Landers, Th. Weber,
H. Schmidt-B\"ocking, R. D\"orner, and T. Jahnke.
\newblock Experimental Proof of Resonant Auger Decay Driven Intermolecular Coulombic Decay.
\newblock {\em Nature, under consideration}, (2012).

\bibitem{Kuleff07prl}
A. I. Kuleff and L. S. Cederbaum.
\newblock Tracing Ultrafast Interatomic Electronic Decay Processes in Real Time and Space.
\newblock {\em Phys. Rev. Lett.}, \textbf{89}, 083201 (2007).

\bibitem{Schollkopf94Science}
W. Sch\"ollkopf, and J. Peter Toennies.
\newblock Nondestructive Mass Selection of Small van der Waals Clusters.
\newblock {\em SCIENCE}, \textbf{266}, 1345 (1994).

\bibitem{Havermeier10prl}
T. Havermeier, T. Jahnke, K. Kreidi, R. Wallauer, S. Voss, M. Sch\"offler, S. Sch\"ossler, L. Foucar, N. Neumann, J. Titze, H. Sann, M. K\"uhnel, J. Voigtsberger, J. H. Morilla, W. Sch\"ollkopf, H. Schmidt-B\"ocking, R. E. Grisenti, and R. D\"orner.
\newblock Interatomic Coulombic Decay following Photoionization of the Helium Dimer: Observation of Vibrational Structure.
\newblock {\em Phys. Rev. Lett.}, \textbf{104}, 133401 (2010).

\bibitem{Sisourat10NatPhys}
N. Sisourat, Nikolai V. Kryzhevoi, P. Kolorenc, S. Scheit, T. Jahnke, and L. S. Cederbaum.
\newblock Ultralong-range energy transfer by interatomic Coulombic decay in an extreme quantum system.
\newblock {\em Nature Physics}, \textbf{6}, 508 (2010).

\bibitem{Scheit03jcp}
S. Scheit, L. S. Cederbaum, and H.-D. Meyer.
\newblock Time-dependent interplay between electron emission and fragmentation in the interatomic Coulombic decay.
\newblock {\em J. Chem. Phys.}, \textbf{118}, 2092 (2003).

\bibitem{Drescher02Nat}
M. Drescher, M. Hentschel, R. Kienberger, M. Uiberacker, V. Yakovlev, A. Scrinzi, Th. Westerwalbesloh, U. Kleineberg, U. Heinzmann, and F. Krausz.
\newblock Time-resolved atomic inner-shell spectroscopy.
\newblock {\em Nature}, \textbf{419}, 803 (2002).

\bibitem{doerner00pr}
R.~D\"orner, V.~Mergel, O.~Jagutzki, L.~Spielberger, J.~Ullrich,
R.~Moshammer, and H.~Schmidt-B\"ocking.
\newblock Cold Target Recoil Ion Momentum Spectroscopy: a 'momentum microscope' to view atomic collision dynamics.
\newblock {\em Physics Reports}, \textbf{330}, 96--192 (2000).

\bibitem{ullrich03rep}
J.~Ullrich, R.~Moshammer, A.~Dorn, R. D\"orner, L.~Ph.~H. Schmidt, and
H.~Schmidt-B\"ocking.
\newblock Recoil-ion and electron momentum spectroscopy: reaction-microscopes.
\newblock {\em Rep. Prog. Phys.}, \textbf{66}, 1463--1545 (2003).

\bibitem{Jahnke04JESRP}
T. Jahnke, Th. Weber, T. Osipov, A. L. Landers, O. Jagutzki, L.
Ph. H. Schmidt, C. L. Cocke, M. H. Prior, H. Schmidt-B\"ocking, and R.
D\"orner.
\newblock Multicoincidence studies of photo and Auger electrons from fixed-in-space molecules using the COLTRIMS technique.
\newblock {\em J. Elec. Spec. Rel. Phen.}, \textbf{73}, 229--238 (2004).

\bibitem{Schuette12prl}
B. Sch\"utte, S. Bauch, U. Fr\"uhling, M. Wieland, M. Gensch, E. Pl\"onjes, T. Gaumnitz, A. Azima, M. Bonitz, and M. Drescher.
\newblock Evidence for Chirped Auger-Electron Emission.
\newblock {\em Phys. Rev. Lett.}, \textbf{108}, 253003 (2012).

\bibitem{Bauch12pra}
S. Bauch, and M. Bonitz.
\newblock Theoretical description of field-assisted postcollision interaction in Auger decay of atoms.
\newblock {\em Phys. Rev.}, \textbf{A85}, 053416 (2012).

\bibitem{Niehaus77jpb}
A. Niehaus.
\newblock Analysis of post-collision interactions in Auger processes following near-threshold inner-shell photoionization.
\newblock {\em J. Phys.}, \textbf{B10}, 1845 (1977).

\bibitem{Sheinerman06jpb}
S. Sheinerman, P. Lablanquie, F. Penent, J. Palaudoux, J. H. D. Eland, T. Aoto, Y. Hikosaka, and K. Ito.
\newblock Electron correlation in Xe 4d Auger decay studied by slow photoelectron-Auger electron coincidence spectroscopy.
\newblock {\em J. Phys.}, \textbf{B39}, 1017 (2006).

\bibitem{Landers09prl}
A. L. Landers, F. Robicheaux, T. Jahnke, M. Sch\"offler, T. Osipov, J. Titze, S.Y. Lee, H. Adaniya, M. Hertlein, P. Ranitovic, I. Bocharova, D. Akoury, A. Bhandary, Th. Weber, M. H. Prior, C. L. Cocke, R. D\"orner, and A. Belkacem.
\newblock  Angular Correlation between Photoelectrons and Auger Electrons from K-Shell Ionization of Neon.
\newblock {\em Phys. Rev. Lett.}, \textbf{102}, 223001 (2009).

\bibitem{Gislason73jcp}
E. A. Gislason.
\newblock Series expansions for Franck-Condon factors. I. Linerar potential and the reflection approximation.
\newblock {\em J. Chem. Phys.}, \textbf{58}, 3702 (1973).

\bibitem{averbukh04prl}
V. Averbukh, I.B. M\"uller, and L.S. Cederbaum.
\newblock Mechanism of Interatomic Coulombic Decay in Clusters.
\newblock {\em Phys. Rev. Lett.}, \textbf{93}, 263002 (2004).

\bibitem{Chiang11prl}
Ying-Chih Chiang, Frank Otto, Hans-Dieter Meyer, and Lorenz S. Cederbaum.
\newblock Interrelation between the Distributions of Kinetic Energy Release and Emitted Electron Energy following the Decay of Electronic States.
\newblock {\em Phys. Rev. Lett.}, \textbf{107}, 173001 (2011).

\bibitem{Kolorenc10pra}
P. Kolorenc, N. V. Kryzhevoi, N. Sisourat, and L. S. Cederbaum.
\newblock Interatomic Coulombic decay in a He dimer: Ab initio potential-energy curves and decay widths.
\newblock {\em Phys. Rev.}, \textbf{A82}, 013422 (2010).

\end{thebibliography}

\end{document}